\documentstyle[11pt,epsfig]{article}

\begin{document}
\begin{center}

\title: {\bf Remarks on ``Adiabatic stabilization: Observation of the
surviving population''}

\author: {Sydney Geltman (JILA) and Mircea
Fotino (MCDB) \\University of Colorado, Boulder, CO 80309-0440,
USA}
\end{center}


\begin{abstract} Questions are raised about certain experimental
and theoretical claims that atoms may be stabilized into their
bound states, and prevented from achieving full ionization, by the
application of adiabatic, ultraintense, high-frequency laser
pulses.  It is pointed out that those authors have used the
weak-field concepts of cross section and ionization rate in an
ultra intense field regime where they have no physical
significance.
\end{abstract}

The purpose of this Comment is to raise several questions
concerning the theoretical interpretation of results obtained from
recent experiments by van Druten et al. [1]. In those measurements
Ne atoms in a cell were prepared in the $ (2p) ^{5} 5g $ $ m=4 $
Rydberg state by a pumping laser and then ionized by a 90 fs pulse
of 620 nm laser radiation having the range of peak intensities
$.05 \times 10^{14}$ to $2.3 \times 10^{14}$ $W/cm^{2}$.  The
relative yields of ions thus produced and the surviving 5g atoms
were measured. Figure 8 in Ref. [1] contains the results of these
relative yields, normalized so they represent the absolute
probabilities P(ion) and P(5g).  Any appreciable leakage into
other channels was assumed negligible, so that the authors
converted their relative measurements to absolute ones with the
normalization ${P(ion) + P(5g)\cong 1}$ at all intensities.  Their
${P(ion) \cong .25}$ and is virtually flat between ${I = .5}
\times 10^{14} W/cm^{2}$ and $2.3 \times 10^{14} W/cm^{2}$, and
the authors point out that this behavior is well below the
ionization probability expected on the basis of the Fermi Golden
Rule.  They also claimed that the relative closeness of their
extracted ionization cross sections to those of the theory of
Potvliege and Smith [2] constituted a verification of the
existence of ``adiabatic stabilization.''

We should like to point out to the reader that a dispute has been
going on over the past decade on whether or not ``adiabatic
stabilization" does indeed exist. Unfortunately, such controversy
was not acknowledged in Ref. [1]. Briefly, the proponents of this
phenomenon claim that when using sufficiently high-intensity,
high-frequency fields, the ionization rate may decrease as the
field intensity increases. Since this is a clearly unexpected and
counter-intuitive behavior, it is not surprising that it is a
matter of considerable theoretical controversy.  For a detailed
presentation of the arguments, we refer the reader to the general
review articles by Eberly and Kulander [3] in favor of
stabilization, and by Geltman [4] in favor of full ionization.
These review articles contain extensive reference lists.

The overall situation as applied to the experiment [1] and theory
[2] under discussion is shown in Fig. 1, where the one-photon
cross section for the ionization of $ Ne ({5g}{m=4})$ is plotted
as a function of laser intensity.  For the purposes of the present
discussion the atom $ Ne ({5g} {m=4})$ may be regarded as purely
hydrogenic.  The 16 points for P(ion) given in Fig. 8 of [1] are
represented here by the points labeled ``exp.'' The theoretical
predictions at high intensity in Fig. 1 are those of Potvliege and
Smith [2] who use a Sturmian-Floquet method to evaluate lifetimes
of various hydrogenic states subject to very intense fields at a
number of wavelengths. In general they find this lifetime to
decrease when the field intensity increases, until a minimum in
lifetime occurs, followed by an increase in lifetime when further
increasing the intensity.  We believe that this unexpected
behavior is most likely the result of the use of a theoretical
method beyond its range of validity.  The Sturmian-Floquet method
cannot account for the time dependence contained in the full
time-dependent Schr\"{o}dinger equation under such extreme
conditions, where the atomic binding potential is effectively
destroyed by the applied field.

\begin{figure}
\caption{Effective cross section of the photoionization of the
hydrogenic ${5g}$  ${m=4}$ state by 620 nm laser radiation as a
function of intensity. The FGR calculated value is indicated at
the lowest intensities (solid line), as are the approximate
regions for the FGR, TI, and OBI modes of ionization. The dashed
line in the TI region is a schematic representation of the cross
section below the OBI threshold (vertical dashed line).
Experimental points of van Druten et al. [1] for $Ne$ ${5g}$
${m=4}$ (filled points) and the calculated values of Potvliege and
Smith [2] (open circles and approximate connecting curve).}
\label{fig1} \epsfbox{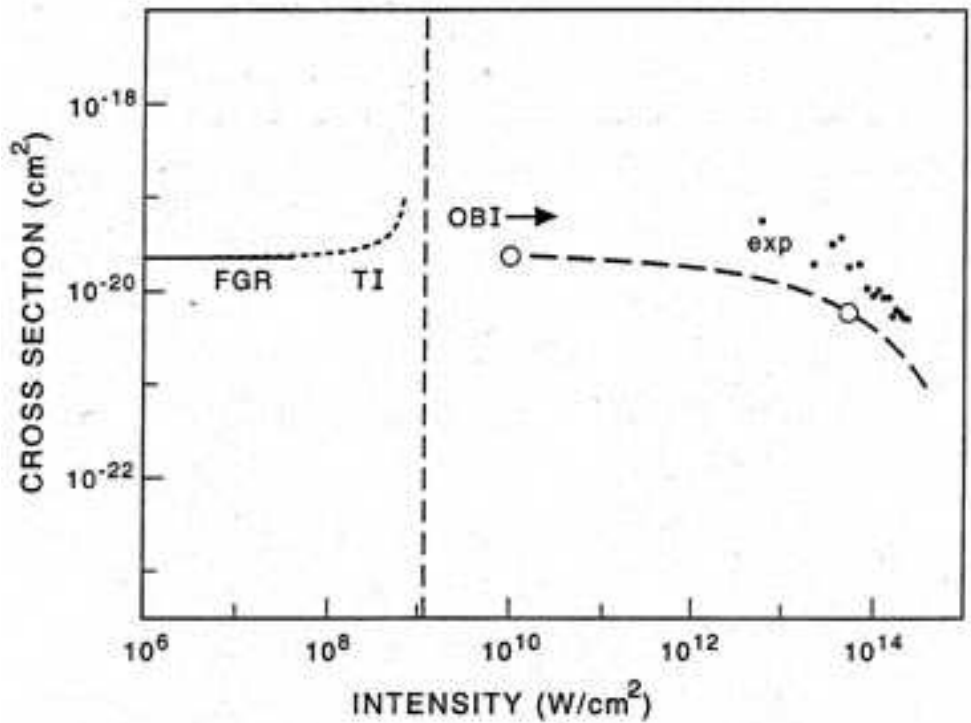}
\end{figure}

It is central to this discussion at this point to indicate the
dilemma that one faces when trying to understand how these results
connect to the picture of photoionization at much lower field
intensities, where there is general agreement.  At the lowest
intensities the Fermi Golden Rule (FGR) allows the evaluation of
exact one-photon photoionization cross sections, and one finds
that ${\sigma}$ $({5g}{m=4}) = 2.15 \times 10^{-20} cm^{2}$ for
${\lambda = 620} {nm}$, using a method of Burgess [5].  This value
is indicated by FGR at the far left of Fig. 1.  At higher
intensities one expects corrections over lowest-order perturbation
theory to enter the picture, and for tunneling (TI) to become the
dominant mechanism for ionization.  At even higher intensities the
``tunnel'' disappears as the effective binding potential falls
below the level of the bound state.  This occurs for a hydrogenic
atom (m=0) when

\begin{equation} {E_o = Z^3/16 n^4},
\end{equation}

where $E_o$ is the peak electric field in the laser pulse (in
a.u.), Z is the effective nuclear charge, and $n$ is the principal
quantum number.  This field strength is the threshold for a very
rapid and nonlinear rise in the probability (to essentially unity)
of the bound electron escaping into the continuum, over the
lowered barrier along the field direction, called over-the-barrier
ionization (OBI).  For the present 5g state and Z=1, one finds
that this condition is reached at a laser intensity of $3.51
\times 10^{8} W/cm^{2}$.  Cooke and Gallagher [6] have pointed out
that when ${m \neq 0}$ there is a correction to the threshold
field value given by (1) which is due to the preservation of the
kinetic energy associated with the angular momentum along the
quantization axis.  The inaccessibility of that energy to the
translational kinetic energy of the ejected electron amounts to a
raising of the effective threshold field of OBI.  Making that
correction we find that the OBI intensity for the ${5g}$ ${m=4}$
state is raised to $1.18 \times 10^{9} W/cm^{2}$.  The above three
regions are indicated by FGR, TI, and OBI in Fig. 1.  The
effective cross section in the TI region has not been evaluated,
but only schematically indicated as the dashed line rising from
the FGR limit to reflect the physically expected increasing cross
section with decreasing tunneling barrier size.

For all intensities above the OBI threshold an effective cross
section may not be meaningfully defined since the ionization
probability is no longer a linear function of the time.  We must
therefore regard as anomalous the experimental [1] and theoretical
[2] points above the OBI threshold in Fig. 1.  In the TI and OBI
regions the more precise quantity to describe the ionization
dynamics is the ionization probability, which results from a
particular laser pulse.  The basic mechanisms for electron
ejection in these regions are tunneling and field emission, which
are qualitatively different from that of photon absorption, the
mechanism that applies at the lowest intensities.  The proper
description in the TI and OBI regions requires the solution of the
full time-dependent Schr\"{o}dinger equation followed by its
projection onto field-free continuum states, a task that has so
far been intractable for real atoms.  Calculations on model atoms
[7] have shown the rapid rise to full ionization, as expected from
the above qualitative arguments, and rigorous deductions of the
absence of stabilization have also been given [8].  The use of an
ionization rate or bound state lifetime, as done in [2], no matter
how sophisticated, can at best describe the true physics only
through the TI region.  Above the OBI threshold the ionization
probability is no longer a linear function of the pulse duration,
and so no ionization rate is any longer physically meaningful.

It is very difficult for a reader who is not actively engaged in
similar experiments to pinpoint exactly where erroneous results
may have arisen in this experiment [1].  An absolute measurement
of ionization probability as presented in Ref. [1] would require
the perfect alignment of three lasers to ensure that they all are
acting on the identical interaction volume in the cell.  For
example, it is not clear that the axial dimensions of all the
focal regions are identical, which would cast doubt on whether all
the $Ne({5g}$ ${m=4})$ atoms produced by the preparation pulse
were exposed to the peak intensity of the ionizing pulse.
Furthermore, the measurements of relative yields of surviving
atoms and ionized atoms may not be reliably converted to absolute
probabilities by the simple normalization used in Fig. 8 of Ref.
[1].  One must have individual absolute probability measurements
for each of these yields to ensure that no other channels are
interfering. Ideally, an accurate absolute measurement of
ionization probability would require the use of crossed atom and
laser beams rather than a gas cell, as it is such crossed beam
geometries that have provided the most accurate measurements in
the past.

In conclusion, the purpose of this Comment is to show the gulf
between a phenomenological understanding of photoionization in
ultra-intense fields and the measurements reported in Ref. [1] and
the theory in Ref. [2].  These reported results on ``adiabatic
stabilization'' use the concept of ionization rates by photon
absorption in an intensity region 3 to 5 orders of magnitude
larger than that at which one would expect the total break-up of
the atom by simple electrostatic arguments.  To become acceptable,
such claims for stabilization must provide a \underline {physical}
answer to the counter argument that the atom is being completely
dissociated at much lower intensities.  How can there be any
appreciable surviving population when the top of the binding
potential barrier lies far below the bound-state energy?  There is
no reason that we can see to expect any appreciable survival
population for a bound atomic state that has been subject to many
($\sim$50) cycles of a field of such extremely high intensity.
\pagebreak

\textbf{REFERENCES}

[1] N. J. van Druten, R. C. Constantinescu, J. M. Schins, H. Nieuwenhuize,
and H. G. Muller, Phys. Rev. A \textbf{55}, 622 (1997).

[2] R. M. Potvliege and P. H. G. Smith, Phys. Rev. A \textbf{48}, R46 (1993).

[3] J. H. Eberly and K. C. Kulander, Science \textbf{262}, 1229 (1993).

[4] S. Geltman, Chem. Phys. Lett. \textbf{237}, 286 (1995).

[5] A. Burgess, Mon. Not. R. Astron. Soc. \textbf{118}, 477 (1958).

[6] W. E. Cooke and T. F. Gallagher, Phys. Rev. A \textbf{17}, 1226 (1978).

[7] Q. Chen and I. B. Bernstein, Phys. Rev. A \textbf{47}, 4099 (1993);
S. Geltman, J. Phys. B \textbf{27}, 1497 (1994).

[8] A. Fring, V. Kostrykin, and R. Schrader, J. Phys. B \textbf{29}, 5651 (1996).

\end{document}